\documentclass[aip,jmp,reprint,onecolumn,showpacs,amsmath,amssymb]{revtex4-1}
\usepackage{graphicx}
\usepackage{bm}
\usepackage[colorlinks=true,linkcolor=black, citecolor=blue, urlcolor=blue,letterpaper,bookmarks=true,bookmarksopen=false,bookmarksnumbered=false,hypertexnames=true]{hyperref}
\usepackage{color}

\newcommand{\pd}[1]{\partial_{#1}}
\newcommand{\bu}{\bm{u}}
\newcommand{\ov}[1]{\overline{#1}}
\newcommand{\wRa}{\widetilde{Ra}}
\newcommand{\eps}{\epsilon}

\newcommand{\llangle}{\left\langle}
\newcommand{\rrangle}{\right\rangle}

\begin{document}
\preprint{Nonlinearity} 
\title{Bounds on Heat Transport in Rapidly Rotating Rayleigh-B\'{e}nard Convection} 
\author{Ian Grooms}\email{grooms@cims.nyu.edu} \affiliation{Courant Institute of Mathematical Sciences, New York University, New York, New York 10012-1185 USA}
\author{Jared P. Whitehead}\affiliation{Department of Mathematics, Brigham Young University, Provo, UT, 84602, USA}

\date{\today} 

\begin{abstract}
The heat transport in rotating Rayleigh-B\'enard convection is considered in the limit of rapid rotation (small Ekman number $E$) and strong thermal forcing (large Rayleigh number $Ra$).
The analysis proceeds from a set of asymptotically reduced equations appropriate for rotationally constrained dynamics; the conjectured range of validity for these equations is $Ra \lesssim E^{-8/5}$.
A rigorous bound on heat transport of $Nu \le 20.56Ra^3E^4$ is derived in the limit of infinite Prandtl number using the background method.
We demonstrate that the exponent in this bound cannot be improved on using a piece-wise monotonic background temperature profile like the one used here.  This is true for finite Prandtl numbers as well, i.e. $Nu \lesssim Ra^3$ is the best upper bound for this particular setup of the background method.  The feature that obstructs the availability of a better bound in this case is the appearance of small-scale thermal plumes emanating from (or entering) the thermal boundary layer.\end{abstract}
\maketitle 

\section{Introduction}
The Rayleigh-B\'enard convection problem is a classical problem in fluid dynamics; it consists of a layer of Boussinesq fluid between cold top and hot bottom boundaries held at constant temperature.
The present investigation addresses the rotating Rayleigh-B\'enard problem, where the system rotates about an axis aligned with gravity.
In this situation the dynamics are described by three nondimensional numbers: the Rayleigh, Ekman, and Prandtl numbers
\[Ra=\frac{g\alpha_T(\Delta\!T)H^3}{\nu\kappa},\quad E=\frac{\nu}{2\Omega H^2},\quad\sigma=\frac{\nu}{\kappa}.\]
The kinematic viscosity is $\nu$, $\kappa$ is the thermal diffusivity, $g$ is the rate of gravitational acceleration, $H$ is the distance between the top and bottom boundaries, $\Omega$ is the system rotation rate, $\alpha_T$ is the thermal expansion coefficient, and $\Delta\!T$ is the magnitude of the temperature difference between the boundaries.
The Taylor number $Ta=E^{-2}$ is sometimes used in place of the Ekman number.

In the limit of rapid rotation (small Ekman number) the dynamics are governed by a set of nondimensional asymptotically reduced equations first derived in Ref.~\onlinecite{JKW98}.
\begin{align} 
\pd{t}w+J[\psi,w]+\pd{z}\psi &=\frac{\wRa}{\sigma}\theta + \nabla_h^2w \tag{1.a}\\
\pd{t}\zeta + J[\psi,\zeta]-\pd{z}w &=\nabla_h^2\zeta\tag{1.b}\\
\pd{t}\theta+J[\psi,\theta]+w\pd{z}\ov{T}&=\frac{1}{\sigma}\nabla_h^2\theta\tag{1.c}\\
E^{-2/3}\pd{t}\ov{T}+\pd{z}\left(\ov{w\theta}\right)&=\frac{1}{\sigma}\pd{z}^2\ov{T}.\tag{1.d}
\end{align}
The Rayleigh number has been rescaled, $\wRa = Ra E^{4/3}$, consistent with the fact that the system becomes linearly stable\cite{Chandra53,ChandraBook} at small Ekman numbers for $Ra\lesssim E^{-4/3}$.
Boundary conditions at $z=0$ and $1$ are $w=\theta=\pd{z}\psi=0$, and $\ov{T}(0)=1$, $\ov{T}(1)=0$.
The vertical velocity is $w$; the horizontal velocity is in geostrophic balance with the pressure $\psi$, which acts as a streamfunction ($u=-\pd{y}\psi$ and $v=\pd{x}\psi$).
Advection is purely horizontal, and is written using the Jacobian operator $J[\psi,\cdot]=\bu\cdot\nabla(\cdot)$. 
The vertical component of vorticity is $\zeta$, which is related to the streamfunction by $\nabla_h^2\psi=\zeta$.
The horizontal coordinates are rescaled to be smaller than the vertical scale (the depth) by a factor of $E^{1/3}$.
The temperature is split into a horizontal mean $\ov{T}$ and a deviation $\theta$, the latter being smaller by a factor of $E^{1/3}$.
The variables $w$, $\psi$, and $\zeta$ all have zero horizontal mean.
In the original derivation\cite{JKW98} the mean temperature $\ov{T}$ evolves on a slower time scale than the other variables; for convenience in this derivation, the time evolution in equation (1.d) uses the same time variable as in equations (1.a-c).
The overbar $\ov{(\cdot)}$ denotes an average over the horizontal coordinates.

Reduced equations for infinite Prandtl number may be derived\cite{SJKW06} from (1.a-d) by rescaling time such that $\pd{t}\to\sigma^{-1}\pd{t}$, rescaling the velocities $\psi\to\sigma^{-1}\psi$, $w\to\sigma^{-1}w$ and then taking $\sigma\to\infty$.
The result is
\begin{align*}
\pd{z}\psi &=\wRa\theta + \nabla_h^2w \tag{2.a}\\
-\pd{z}w &=\nabla_h^2\zeta\tag{2.b}\\
\pd{t}\theta+J[\psi,\theta]+w\pd{z}\ov{T}&=\nabla_h^2\theta\tag{2.c}\\
E^{-2/3}\pd{t}\ov{T}+\pd{z}\left(\ov{w\theta}\right)&=\pd{z}^2\ov{T}.\tag{2.d}
\end{align*}\setcounter{equation}{2}
The same result is reached by first taking the infinite Prandtl number limit of the Boussinesq equations and then taking the small Ekman limit following Ref.~\onlinecite{JKW98}.

Equations (1.a-d) and (2.a-d) are more computationally tractable than their unreduced counterparts, and have been used in computational experiments on rapidly rotating Rayleigh-B\'enard convection\cite{JKW98,SJKW06,JRGK12}.
The asymptotic derivation of the equations\cite{JKW98,SJKW06} suggests that they should be valid for Rayleigh numbers of order $o(E^{-5/3})$; more nuanced analyses\cite{JKRV12,KSB13} suggest that they are actually valid only up to a cutoff of $Ra\sim E^{-8/5}$ or $Ra\sim E^{-3/2}$.\\

The efficiency of convection is measured by the Nusselt number $Nu$, which is the ratio of the total heat transport to the transport that would be affected by conduction alone.
The relationship between the Rayleigh and Nusselt numbers for strong thermal forcing is of perennial interest, and numerical simulations and dimensional scaling arguments lend insight into the relationship\cite{SJKW06,KSNUA09,JRGK12,KSA12}.
These approaches are complemented by upper bound theory, which derives rigorous upper bounds on the Nusselt number from the governing equations, and can sometimes serve to rule out proposed phenomenological scaling relationships\cite{WD11}.
The first rigorous upper bound for the heat transport in Rayleigh-B\'enard convection was derived by Howard\cite{Howard63}, and further extended and analyzed by Busse\cite{Busse69}.
A complementary approach for deriving rigorous upper bounds was developed more recently by Doering and Constantin\cite{DC96}; the relationship between the approaches is explored in Refs.~\onlinecite{DC96,K97,K98}.
Both methods make use of energy integrals, and as a result are unable to take into account the effects of rotation, which does not affect the total energy of the system.
An exception is the case of convection at infinite Prandtl number, where the momentum equations reduce to a linear diagnostic `slaving' relationship whereby the velocity and pressure are determined by the temperature; at infinite Prandtl number rotational effects appear in the slaving relation, which can be used to derive rotation-aware bounds\cite{Chan74,CHP99,DC01,Vitanov03,Vitanov10}.

The present investigation uses the `background' method of Doering and Constantin\cite{DC96} in the context of the reduced equations (1.a-d) and (2.a-d) to derive a rigorous upper bound on the heat transport in rapidly rotating convection.
Section \ref{sec:Pre} presents the fundamentals of the background method for the reduced equations.
Section \ref{sec:Bound} derives a rigorous upper bound of $Nu\le 20.56\wRa^3$ at infinite Prandtl number.
We then demonstrate in section \ref{sec:Opt} that the exponent of $3$ in this bound is optimal for the piece-wise linear background temperature profile used here.
This latter proof is of physical interest because it finds a particular flow configuration that prevents the derivation of a smaller upper bound: small-scale thermal plumes being ejected from the thermal boundary layers.
The optimality condition extends to the case of finite Prandtl number: although we do not derive an upper bound at finite Prandtl number we demonstrate that a bound better than $Nu \le C\wRa^3$ cannot be achieved using the standard background field (i.e.~the exponent of 3 cannot be reduced).
The results are summarized and discussed in section \ref{sec:Disc}.

\section{Preliminaries\label{sec:Pre}}
The variational techniques in upper bound theory require the long-time average of time derivatives to be zero.
Lacking any previously published regularity results for these equations we take this as an assumption, along with the assumption that the Nusselt number
\begin{equation}
Nu = -\limsup_{T\to\infty}\frac{1}{T}\int_0^T\pd{z}\ov{T}(\cdot,t)|_{z=0}\text{d}t\label{eqn:NuDef}
\end{equation}
is well defined.
Thorough consideration of the regularity of solutions to equations (1.a-d) and (2.a-d), an open problem, is beyond the scope of this work.

The background method relies on the use of a specified `background' that consists of a steady solution of the inviscid equations\cite{DC96}; following the usual practice we use a background that consists of a temperature profile $\tau(z)$ that satisfies the boundary conditions on temperature $\tau(1)=0$, and $\tau(0)=1$.
The horizontal mean temperature is thus $\ov{T}=\Theta+\tau$, and (2.c) and (2.d) become
\begin{align}
\pd{t}\theta+J[\psi,\theta]+w\pd{z}\Theta&=\nabla_h^2\theta - w\tau'\label{eqn:theta}\\
E^{-2/3}\pd{t}\Theta+\pd{z}\left(\ov{w\theta}\right)&=\pd{z}^2\Theta + \tau''\label{eqn:Theta}
\end{align}
where the $'$ refers to the derivative with respect to $z$ on $\tau$.

Multiplying (\ref{eqn:theta}) and (\ref{eqn:Theta}) by $\theta$ and $\Theta$ respectively, and averaging over the volume and time leads to the identities
\begin{align}
-\llangle w\theta\pd{z}\Theta\rrangle &= -\llangle|\nabla_h\theta|\rrangle - \llangle\tau'w\theta\rrangle,\text{ and }\\
\llangle w\theta\pd{z}\Theta\rrangle &=-\llangle(\pd{z}\Theta)^2\rrangle + \llangle\Theta\tau''\rrangle.
\end{align}
The bracket notation used here refers to an average over volume and long time.
Adding these leads to
\begin{equation}
 \llangle\Theta\tau''\rrangle = \llangle(\pd{z}\Theta)^2\rrangle+\llangle|\nabla_h\theta|\rrangle + \llangle\tau'w\theta\rrangle.\label{eqn:ThetaTaupp}
\end{equation}
The so-called second power integral for the dynamics, which can be straightforwardly derived from (2.c-d) and (\ref{eqn:NuDef}), relates the Nusselt number to the mean rate of thermal dissipation; it is given by\cite{JRGK12}
\begin{equation}
Nu = \llangle(\pd{z}\ov{T})^2\rrangle + \llangle|\nabla_h\theta|^2\rrangle.
\end{equation}
Using the background decomposition of $\ov{T}$ this becomes
\[Nu = \llangle(\tau')^2\rrangle + \llangle(\pd{z}\Theta)^2\rrangle - 2\llangle\tau''\Theta\rrangle + \llangle|\nabla_h\theta|^2\rrangle,\]
which together with (\ref{eqn:ThetaTaupp}) implies
\begin{equation}
Nu = \llangle(\tau')^2\rrangle - \llangle(\pd{z}\Theta)^2\rrangle -\llangle|\nabla_h\theta|^2\rrangle -2\llangle\tau'w\theta\rrangle.\label{eqn:Nu1}
\end{equation}

The first power integral for the dynamics relates the Nusselt number to the mean rate of viscous dissipation.
The long-time average of equation (2.d), integrated twice in $z$ using the boundary conditions on $\ov{T}$ and the definition of the Nusselt number (\ref{eqn:NuDef}) yields the relation
\begin{equation*}
Nu = 1 + \llangle w\theta\rrangle.
\end{equation*}
Multiplying (2.a) and (2.b) by $w$ and $\psi$ respectively and averaging over the volume and time gives the kinetic energy balance:
\begin{equation*}
\wRa\llangle w\theta\rrangle = \llangle|\nabla_hw|^2\rrangle + \llangle\zeta^2\rrangle.
\end{equation*}
Together these imply the first power integral\cite{JRGK12}
\begin{equation}
Nu = 1 + \frac{1}{\wRa}\left[\llangle|\nabla_hw|^2\rrangle + \llangle\zeta^2\rrangle\right].\label{eqn:Nu2}
\end{equation}

Taking a linear combination of (\ref{eqn:Nu1}) and (\ref{eqn:Nu2}) yields
\begin{equation}
Nu = \frac{1}{1-b}\left(\llangle(\tau')^2\rrangle - b + \mathcal{Q}\right)\label{eqn:Nu3}
\end{equation}
where $b\in[0,1)$ is the `balance parameter' and 
\begin{equation}
\mathcal{Q} = \llangle(\pd{z}\Theta)^2\rrangle +\llangle|\nabla_h\theta|^2\rrangle +\frac{b}{\wRa}\left[\llangle|\nabla_hw|^2\rrangle + \llangle\zeta^2\rrangle\right]+2\llangle\tau'w\theta\rrangle.
\end{equation}
If $b$ and $\tau$ are chosen so as to guarantee that the quadratic form $\mathcal{Q}$ is positive semi-definite for all configurations of $w$, $\theta$, and $\psi$ consistent with the slaving relations (2.a) and (2.b) then (\ref{eqn:Nu3}) implies the bound
\begin{equation}
Nu \le \frac{1}{1-b}\llangle(\tau')^2\rrangle - \frac{b}{1-b}.
\end{equation}
The same steps in the derivation of (\ref{eqn:Nu3}) may be applied to the finite-Prandtl number equations (1.a-d), which leads to the following Prandtl-number dependent quadratic form
\begin{equation}\label{eqn:QFinite}
\mathcal{Q} = \llangle(\pd{z}\Theta)^2\rrangle +\llangle|\nabla_h\theta|^2\rrangle +\frac{b\sigma^2}{\wRa}\left[\llangle|\nabla_hw|^2\rrangle + \llangle\zeta^2\rrangle\right]+2\sigma\llangle\tau'w\theta\rrangle.
\end{equation}
Note that this is equivalent to the quadratic form at infinite Prandtl number after the rescaling $w\to\sigma^{-1}w$, $\zeta\to\sigma^{-1}\zeta$.

In the following we use the piece-wise linear, monotonic background temperature profile, given by
\begin{equation}\label{eqn:tau}
\tau(z)=\left\{\begin{array}{l l}
1-\frac{z}{2\delta} & \text{ for }z\in[0,\delta),\\
\frac{1}{2} & \text{ for }z \in [\delta,1-\delta],\\
\frac{1}{2\delta}(1-z)& \text{ for }z\in(1-\delta,1].\end{array}\right.
\end{equation}
An example of this profile with $\delta=0.2$ is shown in Figure \ref{fig:Optimality}a.

Following  Refs.~\onlinecite{DC01,Yan04,DOR06} we consider the horizontal Fourier transform of the momentum equations (2.a) and (2.b)
\begin{align*}
\pd{z}\hat{\psi} &= \wRa \hat{\theta} -k^2\hat{w},\\
-\pd{z}\hat{w} &= k^4\hat{\psi},
\end{align*}
where $k=|\bm{k}|$ is the modulus of the horizontal wavenumber.
These imply the following slaving relation between the Fourier coefficients of temperature and vertical velocity
\begin{equation}
-\left(\pd{z}^2-k^6\right)\hat{w} = \wRa k^4\hat{\theta}.\label{eqn:Slaving}
\end{equation}

The value of the quadratic form $\mathcal{Q}$ can be related to the horizontal Fourier coefficients of $w$, $\theta$, and $\psi$ by Plancherel:
\begin{equation*}
\mathcal{Q} = \llangle\left(\pd{z}\Theta\right)^2\rrangle+\int \mathcal{Q}_k\text{d}\bm{k}
\end{equation*}
where, for the infinite Prandtl number model, 
\begin{equation}
\mathcal{Q}_{k} = k^2\|\hat{\theta}_{\bm{k}}\|^2 + \frac{b}{\wRa}\left[k^2\|\hat{w}\|^2+\frac{1}{k^4}\|\pd{z}\hat{w}\|^2\right]-\frac{2}{\delta}\mathcal{R}\left\{\int_0^\delta\hat{w}^*\hat{\theta}\text{d}z + \int_{1-\delta}^1\hat{w}^*\hat{\theta}\text{d}z\right\}.\label{eqn:QkInf}
\end{equation}
In the above $\|\cdot\|$ denotes the standard $L^2$ norm in the vertical direction, and $\mathcal{R}\{\cdot\}$ denotes the real part of a complex number.
Clearly, if $\mathcal{Q}_k$ is positive semi-definite for all $\bm{k}$, then $\mathcal{Q}$ will also be positive semi-definite.

In section \ref{sec:Bound} we choose $\delta$ as a function of $\wRa$ such that the term $k^2\|\hat{\theta}\|^2$ dominates the boundary layer integrals (the last terms in (\ref{eqn:QkInf})), making the form positive semi-definite.
The terms involving the balance parameter $b$ do not play a part in the bound, but are considered here to show that the balance parameter will not improve the bound.  This is significant as it also implies that a piece-wise linear, non-monotonic background temperature profile will not improve the bound either (see Ref.~\onlinecite{WhDo2012}).

\section{Upper Bound at Infinite Prandtl Number\label{sec:Bound}}
In this section we derive an upper bound using the Green's function representation of $\hat{w}$ in terms of $\hat{\theta}$, similar to Ref.~\onlinecite{Yan04}.
This leads to a complicated bound on the boundary layer integrals that displays non-uniform behavior in $k$.
We then develop simplified bounds valid for large and small $k$, and show that together they guarantee definiteness of $\mathcal{Q}_k$ for all $k$.

\subsection{A Preliminary Estimate}
The Green's function solution to equation (\ref{eqn:Slaving}) is given by
\begin{equation}
\hat{w}(z) = \frac{k\wRa}{\sinh\left(k^3\right)}\int_0^1g(z,s)\hat{\theta}(s)\text{d}s
\end{equation}
where
\begin{equation}
g(z,s) = \left\{\begin{array}{l l}
\sinh\left(k^3z\right)\sinh\left(k^3(1-s)\right)&\text{ for }z\le s,\\
\sinh\left(k^3s\right)\sinh\left(k^3(1-z)\right)&\text{ for }s\le z.\end{array}\right.
\end{equation}
With this representation, the Cauchy-Schwartz inequality implies the following bound on vertical velocity
\begin{align}\notag
|\hat{w}_k(z)|&\le\frac{k\wRa}{\sinh(k^3)}\|\hat{\theta}_k\|_2\left(\int_0^1g(z,s)^2\text{d}s\right)^{1/2}\\
&=\frac{k\wRa}{\sinh(k^3)}\|\hat{\theta}_k\|_2\left(\int_0^zg(z,s)^2\text{d}s+\int_z^1g(z,s)^2\text{d}s\right)^{1/2}\label{eqn:wCS}\\
&=\frac{\wRa}{2k^{1/2}\sinh(k^3)}\|\hat{\theta}_k\|_2 \sqrt{M(z)}
\end{align}
where
\begin{equation}
M(z) = \left(\sinh \left(2 k^3 z\right)-2 k^3 z\right) \sinh ^2\left(k^3 (1-z)\right)+
\left(2 k^3 (z-1)+\sinh \left(2 k^3 (1-z)\right)\right) \sinh ^2\left(k^3 z\right).
\end{equation}
Inserting into the heat flux and using Cauchy-Schwartz again gives
\begin{align}
\frac{2}{\delta}\mathcal{R}\left\{\int_0^\delta\hat{w}_k^*\hat{\theta}_k\text{d}z\right\}\le\frac{2}{\delta}\int_0^\delta|\hat{w}_k||\hat{\theta}_k|\text{d}z &\le \frac{\wRa}{\delta k^{1/2}\sinh(k^3)}\|\hat{\theta}_k\|_2\int_0^\delta|\hat{\theta}_k| \sqrt{M(z)}\text{d}z\\
&\le\frac{\wRa}{\delta k^{1/2}\sinh(k^3)}\|\hat{\theta}_k\|_2^2\left(\int_0^\delta M(z)\text{d}z\right)^{1/2}\\
&=\wRa\|\hat{\theta}_k\|_2^2\sqrt{N(k,\delta)}
\end{align}
where
\begin{multline}
N(k,\delta) = \frac{\text{csch}^2(k^3)}{2k^4\delta^2}\left(2 \delta  k^6+k^3\left(\delta  \left(\sinh \left(2 k^3\right)+\sinh \left(2 (1-\delta) k^3\right)\right)+(\delta -1) \sinh \left(2 \delta  k^3\right)\right)\right.\\
\left.+\cosh \left(2 (\delta -1) k^3\right)-\cosh \left(2 \delta  k^3\right)-\cosh \left(2 k^3\right)+1\right).\label{eqn:N}
\end{multline}
To bound $N(k,\delta)$ uniformly we first consider the asymptotic limit of large $k$.

\subsection{A Bound for Large $k$}
We prove here that the quadratic form $\mathcal{Q}_k$ is positive semi-definite for $k^3\delta\ge\gamma$ where $\gamma =$arcsinh$(1)\approx0.88$.
Dropping negative terms from (\ref{eqn:N}) and using
\[\sinh\left(2(1-\delta)k^3\right)\le\sinh\left(2k^3\right)\]
implies
\begin{equation}
N(k,\delta) \le \frac{1}{2k^4\delta^2}\left(2 \delta  \frac{k^6}{\sinh^2\left(k^3\right)}+2k^3\delta\frac{\sinh \left(2 k^3\right)}{\sinh^2\left(k^3\right)}+\text{csch}^2\left(k^3\right)\right).
\end{equation}
Since we are dealing with $\delta<1/2$ and $k^3\delta\ge\gamma$ we have the useful inequalities for $k^3\ge2\gamma$
\begin{equation}
\begin{split}
&2\delta<1,\;\;\frac{k^6}{\sinh^2\left(k^3\right)}\le \frac{\gamma^2}{2},\text{ and }\\ &\frac{\sinh\left(2k^3\right)}{\sinh^2\left(k^3\right)}=2\,\text{coth}\left(k^3\right)\le 2\,\text{coth}(2\gamma),\quad \text{ csch}^2\left(k^3\right)\le\text{ csch}^2\left(2\gamma\right).
\end{split}
\end{equation}
These further imply that
\begin{align}
N(k,\delta) &\le \frac{1}{2k^4\delta^2}\left(\frac{\gamma^2}{2}+4k^3\delta \,\text{coth}(2\gamma)+\text{csch}^2(2\gamma)\right),\text{ and }\\
&\le \frac{1}{2k\delta}\left(\frac{\gamma}{2}+4\,\text{coth}(2\gamma)+\frac{\text{csch}^2(2\gamma)}{\gamma}\right).
\end{align}
Since we have a similar estimate near $z=1$, the quadratic form is positive semi-definite for $k$ such that $k^3\delta\ge\gamma$ if
\[k^2 - 2\wRa\left(\frac{1}{2k\delta}\left(\frac{\gamma}{2}+4\,\text{coth}(2\gamma)+\frac{\text{csch}^2(2\gamma)}{\gamma}\right)\right)^{1/2}\ge0,\]
i.e.
\begin{equation}
k^{5/2} \ge \wRa\delta^{-1/2}\left(\gamma+8\,\text{coth}(2\gamma)+\frac{2\,\text{csch}^2(2\gamma)}{\gamma}\right)^{1/2}.\label{eqn:Largek}
\end{equation}
Note, in particular, that if $\delta\sim\wRa^{-3}$ the bound holds for $k\gtrsim\wRa$.

\subsection{A Bound for Small $k$}
To demonstrate $\mathcal{Q}_k$ is positive definite for small $k$, we develop a new bound on $|\hat{w}|$ starting from equation (\ref{eqn:wCS}).
First, note that for $k^3\delta\le \gamma =$arcsinh$(1)$ and $0\le s\le\delta$ we have $\sinh^2\left(k^3s\right)\le \sinh\left(k^3s\right)$.
This implies
\begin{align}\notag
\int_0^z\sinh^2\left(k^3s\right)\sinh^2\left(k^3(1-z)\right)\text{d}s &\le \int_0^z\sinh\left(k^3s\right)\sinh^2\left(k^3(1-z)\right)\text{d}s \\
&= k^{-3}\sinh ^2\left(k^3 (1-z)\right) \left(\cosh \left(k^3 z\right)-1\right),\\\notag
\int_z^1\sinh^2\left(k^3z\right)\sinh^2\left(k^3(1-s)\right)\text{d}s &\le \int_z^1\sinh^2\left(k^3z\right)\sinh\left(k^3(1-s)\right)\text{d}s \\
&=k^{-3}\sinh ^2\left(k^3 z\right) \left(\cosh \left(k^3 (1-z)\right)-1\right),
\end{align}
which lead to the following bound on $|\hat{w}|$
\begin{align}
|\hat{w}_k(z)|&\le\frac{2\wRa}{k^{1/2}\sinh(k^3)}\|\hat{\theta}_k\|_2\sqrt{m(z)}
\end{align}
where
\begin{equation}
m(z) = \sinh ^2\left(\frac{1}{2} k^3 (1-z)\right) \sinh ^2\left(\frac{k^3 z}{2}\right) \left(\cosh \left(k^3 (1-z)\right)+\cosh \left(k^3 z\right)+2\right).
\end{equation}
Elementary considerations further imply that for $0\le z\le \delta$
\begin{equation}
m(z)\le\tilde{m}(z) = \sinh ^2\left(\frac{k^3}{2}\right) \sinh ^2\left(\frac{k^3 z}{2}\right) \left(\cosh \left(k^3\right)+\cosh \left(k^3 \delta\right)+2\right).
\end{equation}
As before this estimate of $|\hat{w}|$ leads to a bound on the boundary layer heat flux of the form
\begin{align}
\frac{2}{\delta}\mathcal{R}\left\{\int_0^\delta\hat{w}_k^*\hat{\theta}_k\text{d}z\right\}\le
\frac{2}{\delta}\int_0^\delta|\hat{w}_k||\hat{\theta}_k|\text{d}z &\le \frac{4\wRa}{\delta k^{1/2}\sinh(k^3)}\|\hat{\theta}_k\|_2\int_0^\delta \sqrt{\tilde{m}(z)}|\hat{\theta}_k|\text{d}z\\
&\le\frac{4\wRa}{\delta k^{1/2}\sinh(k^3)}\|\hat{\theta}_k\|_2^2\left(\int_0^\delta \tilde{m}(z)\text{d}z\right)^{1/2}\\
&=\wRa\,\tilde{n}(k,\delta)
\end{align}
where
\begin{equation}
\tilde{n}(k,\delta) =\frac{\text{sech}\left(\frac{k^3}{2}\right) \sqrt{2\left(\sinh \left(\delta  k^3\right)-\delta  k^3\right) \left(\cosh \left(\delta  k^3\right)+\cosh \left(k^3\right)+2\right)}}{\delta  k^2}.
\end{equation}
We now develop a bound for $k^3\delta\le\gamma$.

Note that $\sinh(k^3\delta)-k^3\delta\le (\sinh(1)-1)(k^3\delta)^3$ for $k^3\delta\in[0,1]$ so that
\begin{align}\notag
\tilde{n}(k,\delta) &\le\frac{\text{sech}\left(\frac{k^3}{2}\right) \sqrt{2\left(k^3\delta\right)^3(\sinh(1)-1)\left(\cosh \left(\delta  k^3\right)+\cosh \left(k^3\right)+2\right)}}{\delta  k^2}\\
&=\left(2k^{5}\delta\right)^{1/2}(\sinh(1)-1)^{1/2}\left(\frac{\cosh \left(\delta  k^3\right)}{\cosh^2\left(\frac{k^3}{2}\right)}+\frac{\cosh \left(k^3\right)+1}{\cosh^2\left(\frac{k^3}{2}\right)} + \text{sech}^2\left(\frac{k^3}{2}\right)\right)^{1/2}.
\end{align}
The following elementary inequalities for $k^3\delta\le\gamma$ and $0<\delta\le1/2$
\begin{equation}
\frac{\cosh \left(\delta  k^3\right)}{\cosh^2\left(\frac{k^3}{2}\right)}\le\frac{\cosh \left(\frac{ k^3}{2}\right)}{\cosh^2\left(\frac{k^3}{2}\right)} = \text{sech}\left(\frac{k^3}{2}\right)\le1,\quad\frac{\cosh(k^3)+1}{\cosh^2(k^3/2)} =2
\end{equation}
imply
\begin{align}
\tilde{n}(k,\delta) \le 2\left(2k^5\delta(\sinh(1)-1)\right)^{1/2}.
\end{align}
The quadratic form $\mathcal{Q}_k$ is positive-definite provided that \begin{equation}
k^2-4(k^5\delta)^{1/2}\wRa\left(2(\sinh(1)-1)\right)^{1/2}\ge0.\label{eqn:Smallk}
\end{equation}
In particular, note that if $\delta\sim\wRa^{-3}$ then the form is positive-definite for $k\lesssim\wRa$.

\subsection{Matching Bounds at Large and Small $k$}
The foregoing analysis bounds the heat flux integrals in the quadratic form $\mathcal{Q}_k$ for $k^3\delta\ge\gamma$ and separately for $k^3\delta\le\gamma$.
These bounds on the convective heat flux in the thermal boundary layers lead to the conditions (\ref{eqn:Largek}) and (\ref{eqn:Smallk}) that guarantee positive-definiteness of $\mathcal{Q}_k$.
Although the bounds on the heat flux integrals are valid for all $k$ regardless of $\delta$, the range of $k$ over which the form is positive-definite depends on the choice of $\delta$.
In this section we choose $\delta$ such that the form is positive-definite for all $\wRa$ and for all $k$.
In particular, we set $\delta=c\wRa^{-3}$.

For large $k$, equation (\ref{eqn:Largek}) implies that the form is positive-definite provided that
\begin{equation}
k \ge \frac{\wRa}{c^{1/5}}\left(\gamma+8\,\text{coth}(2\gamma)+\frac{2\,\text{csch}^2(2\gamma)}{\gamma}\right)^{1/5}.
\end{equation}
To cover the full range of $k^3\delta\ge\gamma$ we must have
\begin{equation}
\frac{\wRa}{c^{1/5}}\left(\gamma+8\,\text{coth}(2\gamma)+\frac{2\,\text{csch}^2(2\gamma)}{\gamma}\right)^{1/5}\le \left(\frac{\gamma}{\delta}\right)^{1/3} = \wRa\left(\frac{\gamma}{c}\right)^{1/3}
\end{equation}
i.e.~
\begin{equation}
c \le \gamma^{5/2}\left(\gamma+8\,\text{coth}(2\gamma)+\frac{2\,\text{csch}^2(2\gamma)}{\gamma}\right)^{-3/2}\approx0.0243.
\end{equation}

For small $k$ equation (\ref{eqn:Smallk}) implies that the form is positive-definite provided that
\begin{equation}
k\le\frac{\wRa}{32 c(\sinh(1)-1)}.
\end{equation}
To cover the full range of $k^3\delta\le\gamma$ we must have
\begin{equation}
\wRa\left(\frac{\gamma}{c}\right)^{1/3}\le\frac{\wRa}{32 c(\sinh(1)-1)}
\end{equation}
i.e.
\begin{equation}
c\le \gamma^{-1/2}(32 (\sinh(1)-1))^{-3/2}\approx 0.0802.
\end{equation}
Since we have not made use of the balance parameter $b$, the least upper bound on $Nu$ is given by setting $b=0$.
Using the stricter condition above,
\begin{equation}
Nu \le \frac{1}{2\delta} = \frac{\wRa^3}{2c}\approx 20.56\wRa^3.
\end{equation}

\section{Optimality\label{sec:Opt}}
\begin{figure}
\begin{center}
\includegraphics[width=\textwidth]{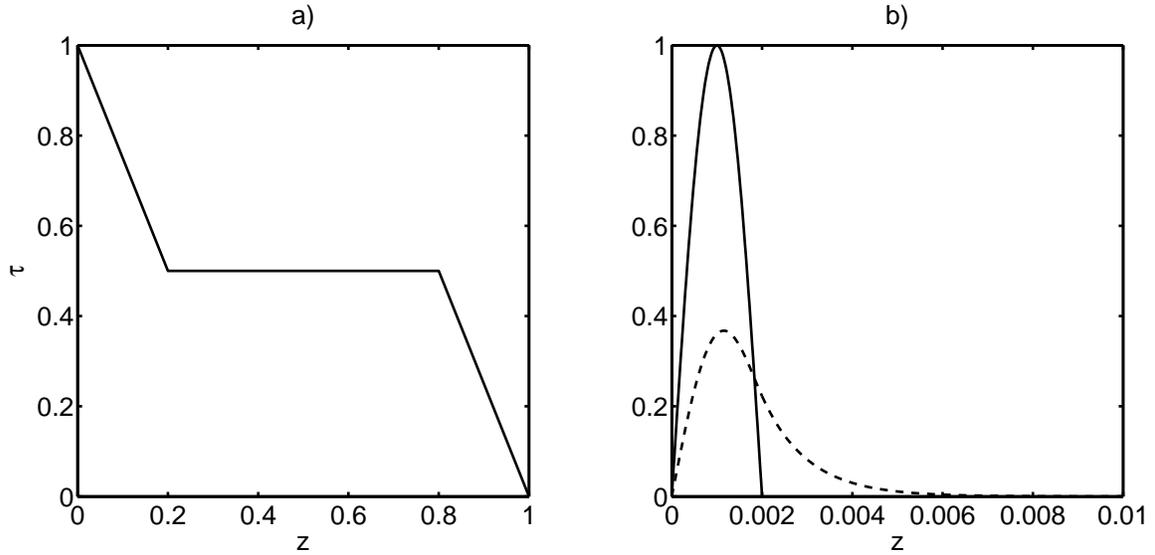}
\end{center}
\caption{\label{fig:Optimality} a) The background profile $\tau$ given by equation (\ref{eqn:tau}) with $\delta=0.2$. b) Profile of temperature perturbation $\hat{\theta}(z)$ (solid) and vertical velocity $\hat{w}(z)$ (dashed) for $k=10$, $\wRa=100$, and $\delta=1/1000$.}
\end{figure}

Although the analysis in the foregoing section provides an upper bound on $Nu$, it is not immediately clear if a more careful analysis might lead to a better bound.
In this section we demonstrate that it will not be possible to achieve a bound $Nu\le C\wRa^\alpha$ with $\alpha<3$ using the standard background temperature profile $\tau$ specified by equation (\ref{eqn:tau}).
The approach is straightforward: we construct fields $\hat{\theta}$ and $\hat{w}$ consistent with the slaving principle (\ref{eqn:Slaving}) such that the quadratic form $\mathcal{Q}_k$ is negative for $\delta\sim\wRa^{-3}$.
We then show that the result extends to finite Prandtl number, i.e.~the test functions also cause the quadratic form to be negative at finite Prandtl number.\\

Consider the flow configuration given by
\begin{equation}
\hat{\theta}_k(z) = \left\{\begin{array}{c l}
\sin\left(\frac{\pi z}{2\delta}\right) & \text{ for }z\in[0,2\delta]\\
0 & \text{ for }z\in(2\delta,1).\\
\end{array}\right.\label{eqn:thetaCE}
\end{equation}
The vertical velocity structure for this configuration is given by the slaving principle (\ref{eqn:Slaving}), which can be integrated analytically to give
\begin{equation}\label{eqn:wCE}
\hat{w}_k(z) = \frac{2k\delta\wRa}{\pi^2+4k^6\delta^2}\left\{\begin{array}{r l}
2k^3\delta\sin\left(\frac{\pi z}{2\delta}\right)+\pi\frac{\sinh\left(k^3(1-2\delta)\right)\sinh\left(k^3z\right)}{\sinh\left(k^3\right)}& \text{ for }z\in[0,2\delta]\\
&\\
\pi\frac{\sinh\left(k^3(1-z)\right)\sinh\left(2k^3\delta\right)}{\sinh\left(k^3\right)}& \text{ for }z\in(2\delta,1].\\
\end{array}\right.
\end{equation}
An example of this flow configuration for $\delta=1/1000$, $k=10$, and $\wRa=100$ is shown in Figure \ref{fig:Optimality}b.
The following analysis would result in similar conclusions for a configuration that is symmetric about $z=1/2$.
A symmetric configuration similar to (\ref{eqn:thetaCE}) is suggested by the structure of $\ov{\theta^2}$ that is observed in simulations\cite{SJKW06,JRGK12}; an antisymmetric profile is chosen for simplicity.

The heat flux integral generated by this configuration is
\begin{equation}
\frac{2}{\delta}\mathcal{R}\left\{\int_0^\delta\hat{w}^*\hat{\theta}\text{d}z\right\} 
=\frac{4k^4\delta^2\wRa}{(\pi^2+4k^6\delta^2)^2}\left(4k^6\delta^2+\pi^2+4\pi\frac{\cosh\left(k^3\delta\right)\sinh\left(k^3(1-2\delta)\right)}{\sinh\left(k^3\right)}\right).\label{eqn:HFCE}
\end{equation}
We seek an example where the quadratic form $\mathcal{Q}_k$ given by equation (\ref{eqn:QkInf}) is negative; guided by the foregoing analysis in section \ref{sec:Bound} we consider small scales $k\sim\wRa$.
The analysis shows that the flow configuration given by equations (\ref{eqn:thetaCE}) and (\ref{eqn:wCE}) leads to a negative quadratic form for large $\wRa$ when $\delta=c\wRa^{-3}$.\\

Let $\delta=c\wRa^{-3}$ and $k=K\wRa$; then the heat flux integral (\ref{eqn:HFCE}) becomes
\begin{equation}
\frac{4c^2K^4}{\wRa(4c^2K^6+\pi^2)^2}\left(4c^2K^6+\pi^2+4\pi\frac{\cosh\left(cK^3\right)\sinh\left(K^3(\wRa^3-2c)\right)}{\sinh\left(K^3\wRa^3\right)}\right).
\end{equation}
The principle part, in the limit $\wRa\to\infty$ is
\begin{equation}
\frac{4c^2K^4}{\wRa(4c^2K^6+\pi^2)^2}\left(4c^2K^6+\pi^2+4\pi\cosh\left(cK^3\right)e^{-2cK^3}\right).
\end{equation}
The thermal dissipation term in $\mathcal{Q}_k$ is simply $k^2\|\hat{\theta}\|^2=\delta k^2=c\wRa^{-1}K^2$.
The viscous dissipation terms in $\mathcal{Q}_k$ are $\mathcal{O}(\wRa^{-4})$, so they can be ignored -- they will not affect the sign of the form in the limit of large $\wRa$.
The form $\mathcal{Q}_k$ thus approaches a negative value in the limit $\wRa\to\infty$ if there are $K$ and $c$ such that
\begin{equation}
cK^2\left(1-\frac{4cK^2\left(4c^2K^6+\pi^2+4\pi e^{-2cK^3}\cosh\left(cK^3\right)\right)}{(4c^2K^6+\pi^2)^2}\right)<0.
\end{equation}
It proves simpler to consider $\xi=cK^3$ and $K$, instead of $c$ and $K$, which results in the following inequality
\begin{equation}
K<\frac{4\xi}{(4\xi^2+\pi^2)^2}\left(4\xi^2+\pi^2+4\pi e^{-2\xi}\cosh\left(\xi\right)\right).
\end{equation}
The right hand side is positive for $\xi>0$ and  achieves a maximum of approximately $0.35$ at $\xi\approx1.224$.
A numerical search indicates that the smallest possible $c$ that allows $\mathcal{Q}_k$ to be negative is approximately $22.5$, and corresponds to $K\approx0.317$.
This indicates that it will not be possible to achieve an upper bound better than about $0.0222\wRa^3$ using a piece-wise linear, monotonic profile at infinite Prandtl number.\\

At finite Prandtl number, equation (\ref{eqn:QFinite}) implies that the quadratic form is
\begin{equation}
\mathcal{Q}_{k} = k^2\|\hat{\theta}_{\bm{k}}\|^2 + \frac{b\sigma^2}{\wRa}\left[k^2\|\hat{w}\|^2+k^4\|\hat{\psi}\|^2\right]-\frac{2\sigma}{\delta}\mathcal{R}\left\{\int_0^\delta\hat{w}^*\hat{\theta}\text{d}z + \int_{1-\delta}^1\hat{w}^*\hat{\theta}\text{d}z\right\}.\label{eqn:QkFinite}
\end{equation}
To derive an upper bound on the Nusselt number at finite Prandtl number this form needs to be positive definite for all possible configurations of $w$, $\theta$, and $\psi$.
The flow configuration given by (\ref{eqn:thetaCE}) and (\ref{eqn:wCE}) can be extended to the finite Prandtl number case by rescaling $w\to \sigma^{-1}w$, and (for simplicity) by taking $\psi=0$.
Thus, using the background $\tau$ specified by equation (\ref{eqn:tau}) it is not possible to derive a bound of $Nu \le C\wRa^\alpha$ with an exponent $\alpha<3$ at any Prandtl number.

\section{Discussion and Conclusions\label{sec:Disc}}
We have applied the background method\cite{DC96} to derive an upper bound on the heat transport in rapidly rotating Rayleigh-B\'enard convection at infinite Prandtl number.
The analysis proceeds from the reduced equations of refs.~\onlinecite{JKW98,SJKW06}, and the main conclusions are that (i) at infinite Prandtl number the Nusselt number is bounded by $Nu \le 20.56Ra^3E^4$, and (ii) it is not possible to derive an upper bound of the form $Nu\le C(RaE^{4/3})^\alpha$ with $\alpha<3$ at any Prandtl number when using the background method with the piece-wise linear, monotonic background temperature profile given by equation (\ref{eqn:tau}).
The regime of validity of this estimate depends on the regime of validity of the reduced equations; they have been conjectured to be valid for $Ra\lesssim E^{-8/5}$ or $Ra\lesssim E^{-3/2}$ (Refs.~\onlinecite{JKRV12,KSB13}, respectively).
Chan\cite{Chan74} also obtained a bound of the form $Nu\lesssim Ra^3E^4$ in rapidly rotating convection at infinite Prandtl number using an asymptotic analysis of the Howard-Busse variational method.

Previous analyses of the unreduced Boussinesq equations at infinite Prandtl number using a variation of the background method have produced the bounds $Nu\le cRa^{2/5}$ independent of $E$ (Ref.~\onlinecite{DC01}), $Nu\le1 + cRa^2 E$ (Ref.~\onlinecite{CHP99}) and $Nu\le cRa^{4/11}(E^{-1}/2+1)^{4/11}$ (Ref.~\onlinecite{Yan04}).
For rapid rotation our bound is the tightest yet; e.g.~for $E= \eps Ra^{-3/4}$ the foregoing upper bounds are $c Ra^{2/5}$, $1+c\eps Ra^{5/4}$, and $c Ra^{4/11}(Ra^{3/4}/(2\eps)+1)^{4/11}$, respectively, while the new upper bound is simply $20.56\eps^4$.

There are two dominant competing phenomenological predictions of the scaling of the Nusselt number with the Rayleigh number in the rapidly rotating regime.
Refs.~\onlinecite{KSNUA09,KSA12} argue following Malkus\cite{Malkus54} that the efficiency of heat flux through the thermal boundary layer controls the net heat flux, and that the thermal boundary layer should remain marginally stable, leading to the prediction $Nu\sim Ra^3E^4$.
In contrast, Refs.~\onlinecite{JRGK12,JKRV12} argue that the fluid interior, away from the boundaries, controls the net heat flux, leading to the alternative scaling law $Nu\sim Ra^{3/2}E^2$, which is also consistent with a so-called `ultimate' regime scaling law that is independent of the viscosity.
Our upper bound is consistent with both of these scaling laws.
Although we show in section \ref{sec:Opt} that there are flow configurations that prevent us from deriving a bound with an exponent smaller than 3, this does not directly support the conjectured behavior of the system for at least two reasons.
First, it may still be possible to derive a smaller upper bound using a different background $\tau$; second, the flow configurations in section \ref{sec:Opt} likely do not appear in the natural evolution of the dynamics.
Indeed, mean temperature $\ov{T}$ in simulations\cite{SJKW06,JRGK12} is significantly different from the $\tau$ (equation \ref{eqn:tau}) used here.

\begin{acknowledgments}
JPW would like to acknowledge discussions with K. Julien and G. Vasil that motivated his interest in this work.
\end{acknowledgments}

\bibliography{GW_Draft_v1}
\end{document}